\documentclass[12pt]{article}
\usepackage[utf8]{inputenc}
\usepackage[T1]{fontenc}
\usepackage{amsmath,amssymb,amsfonts,amsthm}
\usepackage{geometry}
\usepackage{mathpazo}
\usepackage{natbib}
\usepackage{enumerate}
\usepackage{mathrsfs}
\usepackage{booktabs,array,multirow,mathtools,arydshln,threeparttable,caption}
\usepackage{hyperref}
\hypersetup{colorlinks,citecolor=blue,filecolor=black,linkcolor=blue,urlcolor=blue}

\newtheorem{theorem}{Theorem}
\newtheorem{definition}{Definition}
\newtheorem{example}{Example}
\newtheorem{remark}{Remark}
\renewenvironment{proof}[1][Proof]{\emph{#1.} }{\ \hfill $\square$ \vspace{5pt}}

\title{Contractual Structure and Obvious Manipulation\\in Many-to-Many Matching\thanks{We acknowledge the financial support from UNSL through grants 032016 and 030120, from Consejo Nacional de Investigaciones Cient\'ificas y T\'ecnicas (CONICET) through grant PIP 112-200801-00655, and from Agencia Nacional de Promoci\'on Cient\'ifica y Tecnol\'ogica through grant PICT 2017-2355.}}
\author{R. Pablo Arribillaga\thanks{Instituto de Matem\'atica Aplicada San Luis (CONICET--UNSL), Departamento de Matem\'atica, Universidad Nacional de San Luis, San Luis, Argentina, and RedNIE. Email: rarribi@unsl.edu.ar.}\and Beatriz Millan\thanks{Instituto de Matem\'atica Aplicada San Luis (CONICET--UNSL), Departamento de Matem\'atica, Universidad Nacional de San Juan, San Juan, Argentina. Email: millanbetty2@gmail.com.}\and Eliana Pepa Risma\thanks{Instituto de Matem\'atica Aplicada San Luis (CONICET--UNSL), Departamento de Matem\'atica, Universidad Nacional de San Luis, San Luis, Argentina, and RedNIE. Email: ebpepa@unsl.edu.ar.}}
\date{October 6, 2026}
\begin{document}
\maketitle
\begin{abstract}
We study how contractual structure affects worst-case and best-case obvious manipulation in many-to-many matching with substitutable preferences. Efficient individually rational mechanisms rule out best-case obvious manipulations on both the full substitutable domain and the unitary domain. Under simple and economically natural conditions on hospitals' choices, every stable mechanism on the full domain admits obvious manipulations in both cases; hospital-proposing deferred acceptance can admit both types even under unitary doctors' preferences. A further sufficient condition makes efficient stable-dominating mechanisms obviously manipulable in the worst case. On the unitary domain, doctor-proposing deferred acceptance is not obviously manipulable. If each doctor--hospital pair has at most one feasible contract, this positive conclusion extends to every stable-dominating mechanism.
\end{abstract}

\section{Introduction}
Many-to-many matching markets allocate sets of bilateral agreements to agents on both sides of a market. A professional may work for several organizations, a student may receive several training placements, and a buyer may trade with several suppliers. Contractual terms determine not only the identities of partners but also the activities, obligations, or conditions attached to each relationship. Some activities can be agreed separately with the same partner, while other contracts specify mutually exclusive terms. For example, \citet{hatfield2017contract} discuss medical internship arrangements in which surgical and medical placements can be assigned through separate contracts, including placements at the same hospital. Their framework also encompasses buyer--seller markets with heterogeneous indivisible goods. These applications give the many-to-many model an economic role beyond increasing the number of assignments in a many-to-one market.

An agent who receives several contracts must compare entire sets of agreements. The attractiveness of one contract may therefore depend on the contracts the agent can retain elsewhere, even when preferences are substitutable. Representing each position as a separate unit-demand participant would discard these preferences over sets and the fact that one agent controls a single report concerning all its contracts. There is a further distinction within many-to-many markets: several agreements may be signed with the same partner, or the terms of each bilateral relationship may have to be specified in a single contract. The question we address is how these features affect the scope for obvious manipulation of stable mechanisms and of efficient mechanisms that improve upon stable outcomes.

Stability prevents agents from replacing an allocation through mutually acceptable contractual deviations, but it need not deliver an efficient allocation for the side whose preferences are elicited. This motivates considering individually rational allocations that weakly Pareto-dominate a stable allocation for that side. Improving an allocation does not by itself make truthful reporting attractive. We examine stable mechanisms and efficient stable-dominating mechanisms, evaluating welfare for doctors while retaining individual rationality for both doctors and hospitals.

Our incentive criterion is non-obvious manipulability (NOM), introduced by \citet{troyan2020obvious}. A misreport is an obvious manipulation if it improves the worst possible outcome or the best possible outcome relative to truthful reporting, with each comparison evaluated using the agent's true preference. The other agents' reports can vary across the relevant domain. Thus, the criterion compares the extrema of the option sets available under two declarations. It is weaker than strategy-proofness: a NOM mechanism may admit profitable deviations at particular profiles. We keep hospitals' preferences fixed and common knowledge, and study manipulation by doctors who do not know the other doctors' preferences. Our conclusions concern this information structure, rather than uncertainty about hospitals' preferences.

In our setting, the distinction is central to the comparison between stability and efficiency. Contractual alternatives can affect these two comparisons differently, and efficiency places a restriction on one that does not extend to the other. We first show that every efficient individually rational mechanism rules out obvious manipulations in the best case, both with multiple contracts and under unitary preferences. When the other doctors accept no contracts, efficiency gives a doctor its most preferred set among those individually rational for the hospitals. No misreport can generate a better hospital-rational set. This argument does not require stable domination. In matching, the relevant upper bound must additionally respect hospitals' individual rationality, so a doctor cannot generally receive every individually acceptable contract. Stability alone, by contrast, need not attain that upper bound: a doctor-preferred individually rational agreement may be blocked by another acceptable contractual arrangement.

For stable mechanisms in the model with multiple contracts, we identify a simple sufficient condition for best-case obvious manipulation. A hospital regards two contracts with the same doctor as acceptable separately but chooses only one when both are available. We call these \emph{competing contracts}. A doctor willing to sign both may obtain its preferred alternative by declaring the other unacceptable. The condition is local: it does not require that the hospital sign at most one contract in total. If the two alternatives are additionally protected against competition from the rest of the market, the same declaration improves both the worst and best possible outcomes. The same hospital conditions yield best-case obvious manipulation of hospital-proposing deferred acceptance under unitary doctors' preferences, and improve both extrema under protection. We state these conclusions in parallel results, using the examples to illustrate sufficient conditions in fixed markets.

For efficient stable-dominating mechanisms, we identify a different sufficient condition, \emph{contractual replacement with competition}. A doctor has an acceptable contract that can be displaced by contracts with other doctors, together with another contract that the hospital accepts despite competition. By declaring the latter acceptable, the doctor secures an agreement at every profile. Efficiency then forces the mechanism to replace it by the doctor's preferred contract whenever replacement preserves hospital individual rationality. The resulting manipulation improves the worst case, while the general efficiency argument rules out an improvement in the best case. The construction uses unitary reports and therefore applies on both preference domains.

The positive results distinguish a restriction on doctors' preferences from a restriction on the contract set. Under unitary preferences, a doctor never chooses two contracts with the same hospital, although several contractual alternatives may remain available for each pair. Doctor-proposing deferred acceptance is NOM on this domain. If, in addition, each doctor--hospital pair has at most one feasible contract, every stable-dominating mechanism is NOM. Under this restriction on $X$, every substitutable preference is automatically unitary. We therefore treat the stronger conclusion within the unitary analysis, identifying the removal of contractual alternatives as the condition that extends NOM from a particular stable mechanism to the entire stable-dominating class.

Section~\ref{sec:model} presents the model and the two preference domains. Section~\ref{sec:contracts} first studies efficiency and best-case obvious manipulation, then gives sufficient conditions for the negative results, and finally establishes non-obvious manipulability under unitary preferences and a further restriction on the contract set. Section~\ref{sec:final} summarizes the scope of the results and the remaining questions.

\subsection{Relation to the literature}

Our analysis relates to the literature on contractual structure and stability. \citet{hatfield2017contract} show that the language used to express agreements can affect substitutability and the existence of stable outcomes. \citet{kominers2012correspondence} identifies unitarity as a key condition for embedding matching with contracts into matching with salaries. \citet{makominers2018bundling} study incentives to alter the contract language through unilateral bundling. We instead hold the contract set fixed and examine preference misreports, distinguishing the worst-case and best-case comparisons. Our results describe an incentive implication of contractual structure even within domains where substitutability guarantees stable allocations.

The closest work on obvious manipulation in matching is \citet{arribillaga2025obvious}, which studies many-to-one markets with substitutable hospital preferences. The NOM results for doctor-proposing deferred acceptance with contracts and for stable-dominating mechanisms without contracts provide useful benchmarks. Allowing the reporting side to choose sets of contracts raises an additional issue: the treatment of alternatives within a bilateral relationship matters separately from the number of partners. Our analysis makes this issue explicit through conditions for fixed markets and through the contrast between stability and efficiency in the best case. The unit-demand examples isolate the role of contractual alternatives, while the full-domain result for stable mechanisms also uses the willingness to accept several contracts with the same partner.

More broadly, \citet{Blair1988}, \citet{echenique2004theory}, and \citet{klaus2009stable} develop the structure and stability theory of many-to-many matching. Their work also makes clear the need to specify the blocking concept and preference assumptions. On incentives, \citet{hatfield2014many} show that even a stringent max--min preference restriction does not guarantee a stable and strategy-proof mechanism. \citet{chen2016manipulability} relate comparisons of stable mechanisms by manipulability to agents' welfare comparisons. Our question is different: whether a profitable misreport changes an extremal outcome when the other doctors' reports are unknown. NOM can survive the loss of strategy-proofness, but its survival depends on contractual structure and on the class of mechanisms considered.

The stable-domination criterion is studied by \citet{alva2019stabledominating}, and the search for welfare improvements over deferred acceptance is central to \citet{kesten2010school} and \citet{dougan2021minimally}. \citet{alva2019strategy} analyze restrictions on strategy-proof Pareto improvements, while \citet{chen2023regret} study regret-free truth-telling for efficient stable-dominating rules in school choice. A related connection between efficiency and the best-case requirement appears in cake-cutting: \citet[Lemma~4]{ortega2022obvious} show that Pareto-optimal direct-revelation mechanisms satisfy that requirement. \citet{psomas2022fair} also analyze the two comparisons separately in fair allocation of indivisible goods.

Separate analysis of the two comparisons has precedents in several allocation problems. In voting, \citet{aziz2021obvious} distinguish best-case and worst-case obvious manipulation and show that, for positional scoring rules with at least three voters and alternatives, the existence of a best-case obvious manipulation entails the existence of a worst-case obvious manipulation. \citet{arribillaga2024obvious} obtain a related implication for tops-only voting rules. With monetary transfers, \citet{archbold2025singleparameter} show that the two requirements have different implications for subsidies in bilateral trade, and \citet{dekeijzer2026budget} characterize them separately in budget-feasible procurement. In division problems with single-peaked preferences, \citet{article} characterize a broad class of efficient, own-peak-only NOM rules satisfying the equal division guarantee, while \citet{unknown} show that adding consistency singles out the uniform rule. In the same setting, \citet{chen2026disclosure} also considers both the best-case and worst-case requirements when some opponents' reports are disclosed. These results motivate asking how the relationship between the two comparisons depends on the allocation environment.

\section{Preliminaries}\label{sec:model}
\subsection{Matching with contracts, stability, and efficiency}
Let $D$ and $H$ be finite sets of doctors and hospitals, respectively, and define the set of all agents as $F=D\cup H$. There is also a finite set $X$ of feasible contracts, where each contract $x\in X$ represents an agreement between a doctor--hospital pair. Each contract is bilateral, involving exactly one doctor $x_D\in D$ and one hospital $x_H\in H$. The set $X$ may contain two or more contracts involving the same doctor--hospital pair, each specifying different conditions. Given $Y\subseteq X$ and an agent $i\in F$, let $Y_i$ denote the set of all contracts in $Y$ involving $i$, and let $Y_F$ denote the set of agents involved in $Y$.

The matching problem consists of determining which doctors are matched with which hospitals and under what terms; that is, which of the feasible contracts are actually signed. In the general  model (with multiple contracts), each agent can sign multiple contracts with any agent on the opposite side of the market. Therefore, any subset of contracts $Y\subseteq X$ is an \textbf{allocation}, that is, a feasible outcome for the matching problem. 
We denote by $\mathcal{X}$ the set of all allocations.

Each agent $i$ has a strict, complete, transitive preference $P_i$ over subsets of $X_i$, with associated weak preference $R_i$. Choice from an arbitrary offer set $Y\subseteq X$ is
\[
 C_i(P_i,Y)=\max_{P_i}\{Z:Z\subseteq Y_i\}.
\]
We omit $P_i$ when it is understood. 
Throughout the main analysis, preferences are \textbf{substitutable}: for $x\in Y_i\subseteq Z_i$, $x\in C_i(Z)$ implies $x\in C_i(Y)$. The choice satisfies $C_i(Y)\subseteq Y_i$; if $C_i(Y)\subseteq Z\subseteq Y$, then $C_i(Z)=C_i(Y)$; and $C_i(C_i(Y))=C_i(Y)$. Under substitutability, choice is path independent:
\[
 C_i(Y\cup Z)=C_i(C_i(Y)\cup Z).
\]
Write $C_D(Y)=\bigcup_{d\in D}C_d(Y)$ and $C_H(Y)=\bigcup_{h\in H}C_h(Y)$.

An allocation $Y$ is \textbf{individually rational} if $C_D(Y)=C_H(Y)=Y$. A nonempty set $Z\subseteq X\setminus Y$ is a blocking set for $Y$ if $Z_i\subseteq C_i(Y\cup Z)$ for every $i\in Z_F$. An allocation is \textbf{stable} if it is individually rational and has no blocking set. Under substitutability, stability is equivalent to pairwise stability, and stable allocations exist. The doctor-optimal and hospital-optimal stable allocations are obtained by deferred acceptance; see \cite{hatfield2017contract}.

An allocation $Y$ \textbf{Pareto-dominates} $Y'$ for doctors if $Y_dR_dY'_d$ for all $d$ and $Y_dP_dY'_d$ for at least one doctor. All welfare comparisons below concern doctors. An allocation is \textbf{stable-dominating} if it is individually rational and weakly Pareto-dominates some stable allocation, allowing equality. An allocation is \textbf{efficient} if no other individually rational allocation Pareto-dominates it.

\subsection{Mechanisms and obvious manipulations}

In our analysis, we assume that only one side of the market is strategic: the doctors, while hospitals' preferences are fixed and common knowledge. Throughout most of the paper, $P_h$ will be substitutable and fixed, but arbitrary, for each hospital $h$. We will therefore often omit references to hospitals' preferences. 
An \textbf{admissible market} consists of finite sets of doctors $D$, hospitals $H$, and bilateral contracts $X$, together with fixed strict and substitutable hospital preferences.

Within each admissible market, the contracts and hospitals' preferences remain fixed, while doctors' preferences vary over the corresponding domain.

An arbitrary preference for a doctor $d$ is denoted by $P_d$. For each doctor $d$, let $\mathcal{P}_d$ denote the set of all substitutable preferences for $d$. A preference profile $P=(P_d)_{d\in D}$ specifies a preference for each doctor. The set of all substitutable preference profiles is denoted by $\mathcal{P}=\prod_{d\in D}\mathcal{P}_d$. For each profile $P$ and doctor $d$, let $P_{-d}$ denote the subprofile obtained by deleting $P_d$.

In addition to the full substitutable domain, we consider the domain of substitutable unitary preferences. A substitutable preference $P_d$ is \textbf{unitary} if, for every offer set $Y\subseteq X$,
\[
 x,x'\in C_d(P_d,Y),\quad x\ne x'
 \quad\Longrightarrow\quad x_H\ne x'_H.
\]
The terminology follows \cite{kominers2012correspondence}. Let $\mathcal{P}_d^u$ be this domain and $\mathcal{P}^u=\prod_d\mathcal{P}_d^u$. Individual rationality then ensures that no doctor signs two contracts with the same hospital, even though several terms may be feasible for a pair.

Throughout the paper, we work with the full substitutable domain $\mathcal{P}$ and the unitary domain $\mathcal{P}^u$. We use $\mathcal T$ to denote either domain, with individual domains $\mathcal T_d$ and the analogous notation $\mathcal T_{-d}$.

\begin{definition}
A matching mechanism on $\mathcal T$ is a function $\varphi:\mathcal T\to\mathcal{X}$. Write $\varphi_d(P)=\varphi(P)_d$. A mechanism is stable, stable-dominating, or efficient if its output has the corresponding property at every profile.
\end{definition}

We use two stable mechanisms throughout: doctor-proposing deferred acceptance, denoted by $DDAM$, and hospital-proposing deferred acceptance, denoted by $HDAM$. The proposing agents choose among contracts not yet rejected, and the receiving agents choose from accumulated offers. Formally, the doctor-proposing algorithm is as follows.

The doctor-proposing deferred acceptance algorithm starts from $X^1=X$. At iteration $t$, doctors offer $O^t=C_D(X^t)$. Let $O_A^t=\bigcup_{k=1}^tO^k$ be the accumulated offers. Hospitals provisionally accept $C_H(O_A^t)$. If $C_H(O_A^t)=O^t$, the algorithm terminates with that allocation. Otherwise,
\[
 X^{t+1}=X^t\setminus\bigl(O^t\setminus C_H(O_A^t)\bigr),
\]
and the next iteration begins. The resulting mechanism is denoted by ${DDAM}$. Interchanging doctors and hospitals gives the hospital-proposing algorithm and mechanism ${HDAM}$. Both are stable under our assumptions.

We say that $\varphi'$ Pareto-dominates $\varphi$ if $\varphi'_d(P)R_d\varphi_d(P)$ for every doctor and profile. 
\begin{remark}\label{rem:hda}
Every stable-dominating mechanism weakly Pareto-dominates ${HDAM}$. Indeed, every stable allocation is weakly preferred by every doctor to the hospital-optimal stable allocation, and a stable-dominating allocation weakly improves a stable allocation.
\end{remark}

Substitutability alone does not guarantee strategy-proofness in many-to-many matching. We therefore ask which mechanisms avoid obvious manipulations in the sense of \citet{troyan2020obvious}. Hospitals' preferences remain fixed in the following comparisons; only the other doctors' reports vary.

Given $\varphi$, $d$, and a preference $P_d\in\mathcal T_d$, we define the option set left open by $P_d$ at $\varphi$ as follows
\[
 {O}^\varphi(P_d)=\{\varphi_d(P_d,P_{-d}):P_{-d}\in\mathcal T_{-d}\}.
\]
Given a doctor $d$, a preference $P_d$, and a nonempty set of subsets of contracts involving $d$, $\mathcal Y\subseteq\{Y:Y\subseteq X_d\}$, we denote by $W_d(P_d,\mathcal Y)$ and $B_d(P_d,\mathcal Y)$ the worst and best sets of contracts in $\mathcal Y$ according to $P_d$, respectively. When the second argument of $C_d$ is a set of possible sets of contracts, it denotes the most preferred element of that set; when the second argument is a set of contracts, it denotes the choice set defined in Section~2.1. Both comparisons of option sets are evaluated according to the doctor's true preference.

\begin{definition}\label{def:om}
A preference $P'_d$ is a manipulation of $\varphi$ at $P_d$ if
\[
 \varphi_d(P'_d,P_{-d})P_d\varphi_d(P_d,P_{-d})
 \quad\text{for some }P_{-d}\in\mathcal T_{-d}.
\]
It is a \textbf{obvious manipulation in the worst case} if
\begin{equation}\label{eq:wc}
 W_d(P_d,{O}^\varphi(P'_d))P_dW_d(P_d,{O}^\varphi(P_d)),
\end{equation}
and a \textbf{obvious manipulation in the best case} if
\begin{equation}\label{eq:bc}
 B_d(P_d,{O}^\varphi(P'_d))P_dB_d(P_d,{O}^\varphi(P_d)).
\end{equation}
It is an \textbf{obvious manipulation} if at least one of these inequalities holds.
\end{definition}

Observe that each inequality ((\ref{eq:wc}) or (\ref{eq:bc})) implies that $P_d'$ is a manipulation of $P_d$. For~\eqref{eq:wc}, take a subprofile attaining the worst truthful outcome; for~\eqref{eq:bc}, take one attaining the best outcome under the deviation. Finiteness ensures that these extrema are attained.

The following three examples immediately illustrate the distinction. All hospital preferences are substitutable. 

\begin{example}[Only the worst case improves]\label{ex:worst}
Let $D=\{d_1,d_2\}$, $H=\{h\}$, and $X=\{v,w,x\}$, where $v,w$ involve $d_1$ and $x$ involves $d_2$, all at $h$. Let
\[
 P_h:\{v\},\{x\},\{w\},\varnothing.
\]
Let $\varphi$ be any efficient stable-dominating mechanism. Doctor $d_1$'s true preference and deviation are
\[
 P_1:\{w\},\varnothing,
 \qquad P'_1:\{w\},\{v\},\varnothing.
\]
If $d_2$ accepts $x$, the unique stable allocation under $P_1$ is $\{x\}$, and it is efficient. Thus $d_1$ receives $\varnothing$. If $d_2$ rejects $x$, the unique stable allocation is $\{w\}$ (it is efficient). Consequently,
\[
 {O}^\varphi(P_1)=\{\varnothing,\{w\}\}.
\]
Under $P'_1$, $\{v\}$ is the unique stable allocation for either report of $d_2$. Its unique efficient stable-dominating improvement is $\{w\}$: $d_1$ prefers $w$ to $v$, $d_2$ remains unmatched, and hospital individual rationality rules out assigning $x$ together with $w$. Hence
\[
 {O}^\varphi(P'_1)=\{\{w\}\}.
\]
The worst outcome improves from $\varnothing$ to $\{w\}$, while both best outcomes equal $\{w\}$. This is an obvious manipulation in the worst case, but not in the best case. 
\end{example}

\begin{example}[Only the best case improves]\label{ex:best}
Keep the agents and contracts of Example~\ref{ex:worst}, but let
\[
 P_h:\{x\},\{v\},\{w\},\varnothing.
\]
Let $\varphi$ be any stable mechanism. Doctor $d_1$'s true preference and deviation are
\[
 P_1:\{v,w\},\{w\},\{v\},\varnothing,
 \qquad P'_1:\{w\},\varnothing.
\]

If $d_2$ accepts $x$, the unique stable allocation is $\{x\}$ under either report by $d_1$. If $d_2$ rejects $x$, the unique stable allocation is $\{v\}$ under $P_1$ and $\{w\}$ under $P'_1$. In particular, $\{w\}$ is not stable under $P_1$: adding $v$ is attractive to $d_1$, and the hospital prefers $v$ to $w$. Therefore,
\[
 {O}^\varphi(P_1)=\{\varnothing,\{v\}\},
 \qquad
 {O}^\varphi(P'_1)=\{\varnothing,\{w\}\}.
\]
The best outcome improves from $\{v\}$ to $\{w\}$, while both worst outcomes equal $\varnothing$. Thus this is an obvious manipulation in the best case, but not in the worst case.
\end{example}

\begin{example}[Both cases improve]\label{ex:both}
Again use the same agents and contracts, now with
\[
 P_h:\{v\},\{w\},\{x\},\varnothing.
\]
Let $\varphi$ be any stable mechanism. Doctor $d_1$'s true preference and deviation are
\[
 \qquad P_1:\{v,w\},\{w\},\{v\},\varnothing,
 \qquad P'_1:\{w\},\varnothing.
\]
Whatever $d_2$ reports, the unique stable allocation under $P_1$ is $\{v\}$. To see this, $v$ blocks any allocation that does not assign it: $d_1$ accepts $v$ together with $w$ and the hospital ranks $v$ first. Under $P'_1$, the unique stable allocation is $\{w\}$, again independently of $d_2$'s report. Hence
\[
 {O}^\varphi(P_1)=\{\{v\}\},
 \qquad {O}^\varphi(P'_1)=\{\{w\}\}.
\]
Since $\{w\}P_1\{v\}$, the same deviation improves both extrema.
\end{example}

Examples~\ref{ex:best} and~\ref{ex:both} use the doctor's willingness to accept both contracts with the same hospital, an option excluded by unitary preferences. They show why stability alone does not rule out best-case obvious manipulations in the general model. Example~\ref{ex:worst} instead uses unitary preferences and concerns efficient stable-dominating mechanisms. Theorem~\ref{thm:efficient-best} rules out best-case obvious manipulations for efficient individually rational mechanisms on both domains, whereas Theorem~\ref{thm:efficient-worst} identifies a sufficient condition for a worst-case improvement under efficient stable-dominating mechanisms.

\begin{definition}\label{def:nom}
A mechanism $\varphi:\mathcal T\to\mathcal X$ is \textbf{not obviously manipulable} (NOM) if it does not admit obvious manipulations on $\mathcal T$. Otherwise, it is \textbf{obviously manipulable} (OM) on $\mathcal T$. A mechanism is OM in the worst case if it admits an obvious manipulation in the worst case, and OM in the best case if it admits an obvious manipulation in the best case.
\end{definition}
A mechanism may be OM in both the worst case and the best case. This means that it admits manipulations of both types, possibly through different declarations. Example~\ref{ex:both} shows that the same declaration may be an obvious manipulation in both cases.

\section{Main Results}\label{sec:contracts}
The results separate the implications of efficiency, contractual competition, and unitary preferences. We first rule out best-case obvious manipulations for efficient individually rational mechanisms. We then give fixed-market sufficient conditions for obvious manipulation and establish the positive conclusions under unitary preferences.

\subsection{Efficiency and Best-Case Obvious Manipulation}\label{sec:efficiency}
Efficiency places a restriction on the best-case comparison that does not depend on stable domination or on the hospital conditions introduced below. The same argument applies on both preference domains.

\begin{theorem}\label{thm:efficient-best}
In the many-to-many matching model, no efficient individually rational mechanism admits obvious manipulations in the best case. This holds on both the full substitutable domain and the unitary domain.
\end{theorem}
\begin{proof}
Let $\mathcal T$ denote either domain and let $\varphi$ be an efficient individually rational mechanism on $\mathcal T$. Given a doctor $d$ and a true preference $P_d$, define
\[
 \mathcal I_d=\{Y\subseteq X_d:C_H(Y)=Y\},
 \qquad Z=\max_{P_d}\mathcal I_d.
\]
The set $\mathcal I_d$ consists of all sets of contracts involving $d$ that are individually rational for the hospitals. It is nonempty because it contains $\varnothing$. Furthermore, $C_d(P_d,Z)=Z$. Otherwise $C_d(P_d,Z)P_dZ$, and hospital substitutability implies $C_H(C_d(P_d,Z))=C_d(P_d,Z)$, contradicting the definition of $Z$. Thus $Z$ is individually rational for all agents. On the unitary domain, $C_d(P_d,Z)=Z$ also implies that $Z$ is unitary.

Consider the preference subprofile, $P_{-d}$, in which all doctors other than $d$ report empty preferences. Every individually rational allocation at this subprofile belongs to $\mathcal I_d$, and every such allocation different from $Z$ is strictly worse for $d$. All other doctors are unmatched. Consequently, efficiency implies that $\varphi_d(P_d,P_{-d})=Z$ at this subprofile.

On the other hand, for any report $P'_d$ and any subprofile $P'_{-d}$, hospital individual rationality and substitutability imply
\[
 C_H(\varphi_d(P'_d,P'_{-d}))=\varphi_d(P'_d,P'_{-d}).
\]
Indeed, the contracts assigned to $d$ at each hospital form a subset of that hospital's individually rational set of contracts. Hence every possible outcome under the deviation belongs to $\mathcal I_d$ and is weakly worse than $Z$ according to the true preference. Therefore,
\[
 B_d(P_d,{O}^\varphi(P_d))=Z
 R_dB_d(P_d,{O}^\varphi(P'_d))
 \qquad\text{for every }P'_d\in\mathcal T_d.
\]
This rules out obvious manipulations in the best case.
\end{proof}

\subsection{Contractual Competition and Obvious Manipulation}\label{sec:competition}
We now identify configurations of hospital choices that generate obvious manipulations in a fixed admissible market. The first concerns alternative agreements with the same doctor that are acceptable separately but compete when offered together. Protection against other offers strengthens the best-case conclusion to both comparisons.

\begin{definition}\label{def:competing-contracts}\label{def:protected-competing-contracts}
A market admits \textbf{competing contracts} if there exist a doctor $d$, a hospital $h$, and two distinct contracts $v,w\in X_d\cap X_h$ such that
\[
C_h(\{v\})=\{v\},\qquad
C_h(\{w\})=\{w\},\qquad
C_h(\{v,w\})=\{v\}.
\]
Such a pair is \textbf{protected} if, in addition,
\[
v\in C_h(X_h),\qquad
w\in C_h(X_h\setminus\{v\}).
\]
A market admits \textbf{protected competing contracts} if it contains such a protected pair.
\end{definition}

Protection ensures that $v$ is accepted regardless of the remaining offers, while $w$ is accepted whenever $v$ is absent. The next two theorems give parallel conclusions for stable mechanisms on the full domain and for hospital-proposing deferred acceptance on the unitary domain. The hospital conditions are the same, although the doctor's willingness to sign both contracts is used only in the first case.

\begin{theorem}\label{thm:stable-competition}
Consider the many-to-many matching model with multiple contracts.

\medskip

If the market admits competing contracts, every stable mechanism is OM in the best case.

\medskip

If the market admits protected competing contracts, every stable mechanism is OM in both the best and worst cases. Moreover, a single declaration improves both the worst and best possible outcomes.
\end{theorem}

\begin{proof}
\textit{Competing contracts.} Let $d$, $h$, and $v,w$ satisfy the conditions in
Definition~\ref{def:competing-contracts}.
Let $\varphi$ be a stable mechanism. Consider the preference
$P_d\in\mathcal{P}_d$ given by
\[
P_d:\{v,w\},\{w\},\{v\},\varnothing,
\]
with all other sets of contracts ranked below $\varnothing$.
Consider also the alternative declaration
\[
P'_d:\{w\},\varnothing,
\]
with all other sets of contracts ranked below $\varnothing$.
Both preferences are substitutable.

We first show that, for any $P_{-d}\in\mathcal{P}_{-d}$,
no stable allocation at $(P_d,P_{-d})$ assigns $w$ to $d$.
Suppose, to the contrary, that $Y$ is stable and $w\in Y_d$.
Individual rationality for $d$ implies $Y_d\subseteq\{v,w\}$.
Moreover, since $w\notin C_h(\{v,w\})$, substitutability and
individual rationality for $h$ imply $v\notin Y_d$.
Thus $Y_d=\{w\}$.

Again by substitutability,
\[
w\notin C_h(Y\cup\{v\}).
\]
If $v\notin C_h(Y\cup\{v\})$, then
$C_h(Y\cup\{v\})\subseteq Y_h$. By choice consistency,
\[
C_h(Y)=C_h(Y\cup\{v\}),
\]
contradicting individual rationality of $Y$, since
$w\in Y_h=C_h(Y)$. Therefore,
\[
v\in C_h(Y\cup\{v\}).
\]
Since $Y_d=\{w\}$ and $C_d(P_d,\{v,w\})=\{v,w\}$,
doctor $d$ also accepts $v$. Hence $v$ is a blocking contract
for $Y$, a contradiction.

Consequently,
\[
O^\varphi(P_d)\subseteq\{\varnothing,\{v\}\}.
\]
Let $\widehat P_{-d}$ be the subprofile in which every doctor
other than $d$ reports empty preferences. Under
$(P_d,\widehat P_{-d})$, the unique stable allocation is $\{v\}$.
Indeed, $\{v\}$ is individually rational and is not blocked by
$w$, whereas the empty allocation is blocked by $v$.
The preceding argument rules out every allocation containing
$w$. Therefore,
\[
B_d(P_d,O^\varphi(P_d))=\{v\}.
\]

Under $(P'_d,\widehat P_{-d})$, the unique stable allocation is
$\{w\}$. Furthermore, individual rationality under $P'_d$
implies
\[
O^\varphi(P'_d)\subseteq\{\varnothing,\{w\}\}.
\]
Thus,
\[
B_d(P_d,O^\varphi(P'_d))=\{w\}.
\]
Since $\{w\}P_d\{v\}$, we obtain
\[
B_d(P_d,O^\varphi(P'_d))
P_d
B_d(P_d,O^\varphi(P_d)).
\]
Hence $P'_d$ is an obvious manipulation in the best case.

\textit{Protected competing contracts.} Use the same preferences $P_d$ and $P'_d$ as in the competing-contract argument, now under the protection condition. For any $P_{-d}$, the preceding argument shows that a stable allocation under $P_d$ assigns either $\varnothing$ or $\{v\}$ to $d$. The empty assignment is impossible: substitutability and $v\in C_h(X_h)$ imply that $h$ accepts $v$ from any offer set containing it, so $v$ would block. Thus $d$ always receives $\{v\}$.

Under $P'_d$, individual rationality excludes $v$ and every contract for $d$ other than $w$. If $d$ is unmatched in a stable allocation $Y$, then $Y_h\cup\{w\}\subseteq X_h\setminus\{v\}$. Substitutability and $w\in C_h(X_h\setminus\{v\})$ imply that $w$ blocks $Y$. Consequently,
\[
O^\varphi(P_d)=\{\{v\}\},\qquad
O^\varphi(P'_d)=\{\{w\}\}.
\]
Since $\{w\}P_d\{v\}$, the same declaration improves both extrema.

\end{proof}

\begin{theorem}\label{thm:hdam-competition}
Consider the many-to-many matching model with unitary contracts.

\medskip

If the market admits competing contracts, $HDAM$ is OM in the best case.

\medskip

If the market admits protected competing contracts, $HDAM$ is OM in both the best and worst cases. Moreover, a single declaration improves both the worst and best possible outcomes.
\end{theorem}

\begin{proof}
Adapt the proof of Theorem~\ref{thm:stable-competition} by using
\[
P_d:\{w\},\{v\},\varnothing,
\qquad P'_d:\{w\},\varnothing,
\]
with every other set below $\varnothing$. Under $P_d$, $HDAM$ never offers $w$: before its first offer of $w$, doctor $d$ cannot have rejected $v$, whereas substitutability excludes choosing $w$ while $v$ remains available. Thus the empty-preference subprofiles yield the same best outcomes as in the competing-contract argument. Under protection, $v$ is offered initially and retained under $P_d$, while the preceding protection argument forces $w$ under $P'_d$. The same singleton option sets follow, proving both conclusions for $HDAM$.
\end{proof}

For efficient stable-dominating mechanisms, we use a different configuration. Competition can exclude a doctor's preferred contract $w$, while another contract $v$ is accepted regardless of the remaining offers. The replacement requirement ensures that replacing $v$ by $w$ preserves hospital individual rationality and leaves the contracts with other doctors unchanged.

\begin{definition}\label{def:contractual-replacement}
A market admits \textbf{contractual replacement with competition}
if there exist a doctor $d$, a hospital $h$, two distinct contracts
$v,w\in X_d\cap X_h$, and a set $A\subseteq X_h\setminus X_d$,
containing at most one contract with each doctor, such that
\[
v\in C_h(X_h),\qquad
C_h(\{w\})=\{w\},\qquad
C_h(A\cup\{w\})=A,
\]
and, for every $Y_h\subseteq X_h$ satisfying
$C_h(Y_h)=Y_h$, $v\in Y_h$, and $w\notin Y_h$,
\[
C_h\bigl((Y_h\setminus\{v\})\cup\{w\}\bigr)
=(Y_h\setminus\{v\})\cup\{w\}.
\]
\end{definition}

\begin{theorem}\label{thm:efficient-worst}
In the many-to-many matching model, if the market admits contractual replacement with competition, any efficient stable-dominating mechanism is OM in the worst case. This holds on both the full substitutable domain and the unitary domain.
\end{theorem}
\begin{proof}
Let $\mathcal T$ denote either domain.

Let $d,h,v,w,A$ satisfy
Definition~\ref{def:contractual-replacement}, and let $\varphi$
be an efficient stable-dominating mechanism. Consider
\[
P_d:\{w\},\varnothing,
\qquad
P'_d:\{w\},\{v\},\varnothing,
\]
with all other sets of contracts ranked below $\varnothing$.
Both preferences are substitutable and unitary.

Let $\widehat P_{-d}$ be the subprofile in which each doctor
involved in a contract of $A$ accepts only that contract, and
all remaining doctors report empty preferences. This subprofile
belongs to both domains. We first show that $A$ is the unique
stable allocation at $(P_d,\widehat P_{-d})$.

Individual rationality restricts every allocation to a subset
of $A\cup\{w\}$. Moreover, substitutability and
$C_h(A\cup\{w\})=A$ imply $C_h(A)=A$, so $A$ is individually
rational. It is not blocked by $w$, since $h$ rejects $w$ from
$A\cup\{w\}$. All other contracts are unacceptable to their
doctors. Thus $A$ is stable.

Conversely, if an individually rational allocation $Y$ omits
some $a\in A$, then
\[
a\in C_h(Y\cup\{a\}),
\]
by substitutability and $a\in C_h(A\cup\{w\})$. The doctor
involved in $a$ is unmatched and accepts $a$, so $a$ blocks
$Y$. Every stable allocation therefore contains $A$.
Hospital individual rationality excludes $A\cup\{w\}$,
establishing uniqueness.

Stable domination consequently requires $\varphi$ to retain
every contract of $A$. Individual rationality excludes all
contracts outside $A\cup\{w\}$ and prevents adding $w$ to $A$.
Hence
\[
\varphi_d(P_d,\widehat P_{-d})=\varnothing.
\]
Since individual rationality under $P_d$ restricts the doctor's
outcomes to $\varnothing$ and $\{w\}$,
\[
W_d(P_d,O^\varphi(P_d))=\varnothing.
\]

Now fix any subprofile $P_{-d}$ in the relevant domain.
Every stable allocation $Y$ at $(P'_d,P_{-d})$ assigns either
$v$ or $w$ to $d$. Indeed, individual rationality permits only
$\varnothing$, $\{v\}$, and $\{w\}$. If $Y_d=\varnothing$,
then substitutability and $v\in C_h(X_h)$ imply
\[
v\in C_h(Y\cup\{v\}).
\]
Since $d$ also accepts $v$, this contract blocks $Y$.
Stable domination and individual rationality therefore imply
\[
\varphi_d(P'_d,P_{-d})\in\{\{v\},\{w\}\}.
\]

Suppose that $\varphi_d(P'_d,P_{-d})=\{v\}$, and write
$Z=\varphi(P'_d,P_{-d})$. Individual rationality gives
$C_h(Z_h)=Z_h$, while $v\in Z_h$ and $w\notin Z_h$.
The replacement condition implies
\[
C_h\bigl((Z_h\setminus\{v\})\cup\{w\}\bigr)
=(Z_h\setminus\{v\})\cup\{w\}.
\]
Consequently, $(Z\setminus\{v\})\cup\{w\}$ is individually
rational for all agents. It makes $d$ strictly better off
according to $P'_d$ and leaves every other doctor's assignment
unchanged, contradicting efficiency.

Thus $\varphi_d(P'_d,P_{-d})=\{w\}$ for every $P_{-d}$, and
\[
O^\varphi(P'_d)=\{\{w\}\}.
\]
It follows that
\[
W_d(P_d,O^\varphi(P'_d))
=\{w\}
P_d
\varnothing
=W_d(P_d,O^\varphi(P_d)).
\]
Hence $P'_d$ is an obvious manipulation in the worst case.
The argument applies on both domains.
\end{proof}

\begin{remark}\label{rem:hospital-conditions}
The three conditions describe simple configurations of hospital choices and do not require unit demand for the hospital as a whole. Competing contracts generalize Examples~\ref{ex:best} and~\ref{ex:both}: let the hospital choose at most one acceptable contract from a group containing $v,w$, ranking $v$ above $w$, and choose independently from its remaining contracts according to any substitutable preference. The same pair is protected if $v$ and $w$ are the first and second choices in the group, as in Example~\ref{ex:both}.

For contractual replacement with competition, generalizing Example~\ref{ex:worst}, let the hospital choose at most one contract from a group $\{v,x,w\}$, with all three acceptable and ranked in that order, where $x$ involves another doctor. Let it choose independently from the remaining contracts according to any substitutable preference. Then $A=\{x\}$ excludes $w$, $v$ is accepted regardless of other offers, and replacing $v$ by $w$ preserves individual rationality. In all three constructions the hospital may simultaneously sign several contracts for other positions or activities, and the other hospitals' preferences remain unrestricted apart from substitutability. The conditions are sufficient, rather than necessary, for the corresponding manipulations.
\end{remark}

Examples~\ref{ex:both} and~\ref{ex:worst}, together with Theorems~\ref{thm:stable-competition}, \ref{thm:hdam-competition}, and~\ref{thm:efficient-worst}, show that neither stable mechanisms on the full domain, nor $HDAM$ on the unitary domain, nor efficient stable-dominating mechanisms on either domain can be NOM in all admissible markets. 

\subsection{Non-Obvious Manipulability under Unitary Preferences}\label{sec:unitary}
Recall that $\mathcal P^u=\prod_d\mathcal P_d^u$ is the domain of substitutable unitary doctors' preferences. On this domain, doctors never choose two contracts with the same hospital, although several contractual alternatives may remain available for each pair. Theorem~\ref{thm:hdam-competition} shows that $HDAM$ can still admit both types of manipulation. Doctor-proposing deferred acceptance has a different implication.

\begin{theorem}\label{thm:unitary}
In the many-to-many matching model with unitary contracts, $DDAM$ is NOM in every admissible market.
\end{theorem}
\begin{proof}
Fix a doctor $d$ and a true preference $P_d\in\mathcal P_d^u$. Write $X_{ih}=X_i\cap X_h$. We first establish a bound on the worst truthful outcome and then construct an admissible subprofile that bounds the outcome of every alternative report.

We use the following elementary property of unitary substitutable preferences. If $x\in C_i(P_i,Y)$ and $x_H=h$, then
\begin{equation}\label{eq:local-choice}
 C_i(P_i,Y_h)=\{x\}.
\end{equation}
Indeed, substitutability implies that $x$ is chosen from $Y_h$, and unitarity excludes any other contract with $h$. Thus, whenever a doctor chooses a contract with $h$, it is the most preferred available singleton at $h$, taking $\varnothing$ into account.

We also use the DA identity
\begin{equation}\label{eq:memory}
 C_H(O_A^t)=C_H(O^t).
\end{equation}
For completeness, the equality holds at $t=1$ because $O_A^1=O^1$. Suppose it holds at $t-1$. Every contract in $C_H(O_A^{t-1})$ is then offered at iteration $t-1$ and is not rejected then. By substitutability of doctors' preferences it is offered again at $t$. Hence $C_H(O_A^{t-1})\subseteq O^t$, and path independence gives
\[
 C_H(O_A^t)
 =C_H\bigl(C_H(O_A^{t-1})\cup O^t\bigr)
 =C_H(O^t).
\]
This proves the identity by induction. Rejections are permanent by substitutability. Consequently, if $A^*$ is the accumulated offer set at termination and $R^*$ is the set of rejected contracts, then
\begin{equation}\label{eq:rejected}
 R^*=A^*\setminus C_H(A^*).
\end{equation}

\medskip
\noindent\textbf{Claim 1.}
Suppose two profiles have substitutable doctor choices $C_i$ and $\overline C_i$ satisfying
\[
 C_i(Y)\subseteq\overline C_i(Y)
 \qquad\text{for every doctor }i\text{ and offer set }Y.
\]
With hospitals' choices unchanged, every contract rejected by DA under $C_D$ is also rejected by DA under $\overline C_D$.

\smallskip
\noindent\emph{Proof of Claim 1.}
Let $\overline A$ and $\overline R$ be the final accumulated offer set and rejection set under $\overline C_D$, and let $\overline X=X\setminus\overline R$. The final offer set is $\overline C_D(\overline X)=C_H(\overline A)$.

We proceed by induction over iterations of DA under $C_D$. At the first iteration,
\[
O^1=C_D(X)\subseteq\overline C_D(X)\subseteq\overline A.
\]
If $x$ is rejected at this iteration, then $x\notin C_H(O^1)$. Since $O^1\subseteq\overline A$, substitutability implies $x\notin C_H(\overline A)$, and hence $x\in\overline R$ by~\eqref{eq:rejected}.

For the inductive step, suppose all contracts rejected before iteration $t$ belong to $\overline R$, so that $\overline X\subseteq X^t$. If $x\in C_i(X^t)$ and $x\notin\overline R$, substitutability implies
\[
 x\in C_i(\overline X)\subseteq\overline C_i(\overline X)
 \subseteq\overline A.
\]
If instead $x\in\overline R$, it also belongs to $\overline A$. Thus $O^t\subseteq\overline A$, and the same holds for $O_A^t$. If $x$ is rejected at $t$, then $x\notin C_H(O_A^t)$. Substitutability implies $x\notin C_H(\overline A)$, and therefore $x\in\overline R$ by~\eqref{eq:rejected}. This completes the induction. \hfill$\square$

\medskip
For each $h$, consider the auxiliary market consisting of hospital $h$, the doctors, and the contracts $X_h$. Every doctor in this auxiliary market has unit-demand preferences: it can sign at most one contract. Hospital $h$ retains its original preference. Doctor $d$ ranks singletons and $\varnothing$ as under $P_d$; denote this order by $P_d^h$. The other doctors' strict unit-demand orders vary freely. Let $V_h$ be the worst outcome of $d$ under doctor-proposing DA in this auxiliary market, evaluated using $P_d^h$. Thus $V_h$ is empty or a singleton. Define
\begin{equation}\label{eq:V}
 V=\bigcup_{h\in H}V_h,
 \qquad L=C_d(P_d,V).
\end{equation}
Finiteness ensures that each $V_h$ is attained.

\medskip
\noindent\textbf{Claim 2.}
For every $P_{-d}\in\mathcal P_{-d}^u$,
\begin{equation}\label{eq:truth-bound}
 {DDAM}_d(P_d,P_{-d})\;R_d\;L.
\end{equation}

\smallskip
\noindent\emph{Proof of Claim 2.}
At an arbitrary profile $P=(P_d,P_{-d})$, complete every doctor's choice by making its decisions at different hospitals independent:
\begin{equation}\label{eq:completion}
 \overline C_i(Y)=\bigcup_{h\in H}C_i(P_i,Y_h).
\end{equation}
These completed choices are substitutable and unitary: at each hospital they choose the best acceptable singleton according to $P_i$, or nothing if none is acceptable. They can be induced by strict preferences for the doctors that evaluate contracts independently across hospitals. To construct such preferences, assign contract values preserving the original order of singletons and $\varnothing$ at each hospital, add these values on unitary sets, and place nonunitary sets below $\varnothing$. The values can be chosen to avoid ties. Hospitals' preferences remain unchanged. Moreover, by substitutability,
\[
 C_i(P_i,Y)\subseteq\overline C_i(Y).
\]

Let $\overline S$ be the DA outcome under the completed choices. This expanded market decomposes into the auxiliary single-hospital markets. The completion preserves each doctor's choices from contracts with a single hospital; in particular, it preserves the order $P_d^h$ of $d$ over singletons and $\varnothing$. Thus $\overline S_{dh}$ is one of the auxiliary outcomes over which $V_h$ was defined, and
\[
 \overline S_{dh}\;R_d\;V_h.
\]
Both sets in this comparison are empty or singletons. If $v\in V_h\setminus\overline S_{dh}$, then $\overline S_{dh}=\{s\}$ for some $s$ with $\{s\}P_d\{v\}$. Unitarity implies $C_d(P_d,\{s,v\})=\{s\}$, and substitutability therefore excludes $v$ from $C_d(P_d,\overline S_d\cup V)$. Consequently,
\[
 C_d(P_d,\overline S_d\cup V)\subseteq\overline S_d.
\]
Choice consistency then gives
\[
 C_d(P_d,\overline S_d\cup V)=C_d(P_d,\overline S_d),
\]
and hence
\begin{equation}\label{eq:S-bound}
 C_d(P_d,\overline S_d)\;R_d\;C_d(P_d,V)=L.
\end{equation}

By Claim 1, completion enlarges the rejection set. In particular, no contract in $\overline S_d$ was rejected in the original run at $P$. If $X^*$ is the final available set in that original run, then $\overline S_d\subseteq X^*$, and therefore
\[
 {DDAM}_d(P)=C_d(P_d,X^*)\;R_d\;C_d(P_d,\overline S_d).
\]
Together with~\eqref{eq:S-bound}, this proves~\eqref{eq:truth-bound}. \hfill$\square$

\medskip
\noindent\textbf{Claim 3.
}
For every hospital $h$, there is a set $B_h\subseteq X_h\setminus X_d$, containing at most one contract with each other doctor, such that $C_h(B_h)=B_h$ and
\begin{equation}\label{eq:blocker}
 \{x\}P_d V_h
 \quad\Longrightarrow\quad
 C_h(B_h\cup\{x\})=B_h
 \qquad (x\in X_{dh}).
\end{equation}

\smallskip
\noindent\emph{Proof of Claim 3.}
If no singleton is strictly preferred to $V_h$, take $B_h=\varnothing$. Otherwise, choose an auxiliary profile attaining $V_h$ and run doctor-proposing DA. Let $q$ be the least preferred contract among those whose singletons are strictly preferred to $V_h$. All these contracts are acceptable, since the truthful outcome $V_h$ is weakly preferred to $\varnothing$. They are offered in descending order and rejected before the doctor obtains $V_h$.

Let $k$ be the iteration in which $q$ is rejected, and put
\[
 B_h=C_h(O_A^k).
\]
At this iteration $q$ is the only offer of $d$. By~\eqref{eq:memory}, $B_h=C_h(O^k)$, so $B_h$ contains no contract of $d$ and at most one contract with each other doctor. Idempotence implies $C_h(B_h)=B_h$.

Every $x$ with $\{x\}P_dV_h$ belongs to $O_A^k$, while $B_h$ contains no contract of $d$. Thus
\[
 B_h\subseteq B_h\cup\{x\}\subseteq O_A^k.
\]
Choice consistency gives $C_h(B_h\cup\{x\})=C_h(O_A^k)=B_h$, proving~\eqref{eq:blocker}. \hfill$\square$

\medskip
We now combine the sets in Claim 3. Put $B=\bigcup_h B_h$. For each $i\ne d$, choose a strict preference $\widetilde P_i$ whose choice is
\begin{equation}\label{eq:rivals}
 C_i(\widetilde P_i,Y)=Y_i\cap B_i.
\end{equation}
Such a preference can be obtained by assigning strictly positive generic additive values to contracts in $B_i$ and placing every set containing a contract outside $B_i$ below $\varnothing$. It is substitutable and unitary because $B_i$ contains at most one contract with each hospital. All contracts in $B$ are offered initially, and no contract outside $B$ is ever offered by the other doctors.

Fix any report $P'_d\in\mathcal P_d^u$ and let
\[
 Y={DDAM}_d(P'_d,\widetilde P_{-d}).
\]
Every $x\in Y_h$ is accepted from an accumulated offer set containing $B_h\cup\{x\}$. Substitutability therefore implies
\[
 x\in C_h(B_h\cup\{x\}).
\]
By~\eqref{eq:blocker}, it follows that
\begin{equation}\label{eq:local-upper}
 V_h\;R_d\;\{x\}
 \qquad\text{for every }x\in Y_h.
\end{equation}

We claim that $C_d(P_d,V\cup Y)\subseteq V$. Indeed, take $x\in Y\setminus V$ and let $h=x_H$. If $V_h=\varnothing$, then~\eqref{eq:local-upper} implies $\varnothing P_d\{x\}$, so substitutability excludes $x$ from every choice containing it. If $V_h=\{v\}$, strictness and~\eqref{eq:local-upper} imply $\{v\}P_d\{x\}$; since $v$ is acceptable and preferences are unitary, $C_d(P_d,\{v,x\})=\{v\}$. Substitutability again excludes $x$ from $C_d(P_d,V\cup Y)$. By choice consistency,
\[
 C_d(P_d,V\cup Y)=C_d(P_d,V)=L.
\]
As $Y\subseteq V\cup Y$, this yields
\begin{equation}\label{eq:deviation-bound}
 L\;R_d\;{DDAM}_d(P'_d,\widetilde P_{-d})
 \qquad\text{for every }P'_d\in\mathcal P_d^u.
\end{equation}
Claims 2 and~\eqref{eq:deviation-bound}, applied also to $P'_d=P_d$, show that
\[
 W_d(P_d,O^{{DDAM}}(P_d))=L
 ={DDAM}_d(P_d,\widetilde P_{-d}).
\]
For any alternative report,
\[
\begin{aligned}
 W_d(P_d,O^{{DDAM}}(P_d))
 &=L\\
 &R_d\;{DDAM}_d(P'_d,\widetilde P_{-d})\\
 &R_d\;W_d(P_d,O^{{DDAM}}(P'_d)).
\end{aligned}
\]
Thus no obvious manipulation in the worst case exists.

Finally, let
\[
 E_d=\{x\in X_d:\{x\}P_{x_H}\varnothing\},
 \qquad Z=C_d(P_d,E_d).
\]
When all other doctors accept no contracts, every offer by $d$ is the only current offer at its hospital. Hence hospital-acceptable offers are accepted and hospital-unacceptable offers are rejected. At termination the available set contains $E_d$ and the doctor's outcome is contained in $E_d$; choice consistency implies that this outcome equals $Z$. Conversely, individual rationality for hospitals and substitutability imply that the outcome of $d$ under any report and any subprofile is a subset of $E_d$. No such set is preferred to $Z$. Therefore, the best truthful outcome cannot be improved either, and ${DDAM}$ is NOM.
\end{proof}

Theorem~\ref{thm:unitary} does not establish the uniqueness of $DDAM$ among NOM stable mechanisms on the unitary domain. A stronger positive conclusion follows from restricting the contractual structure itself: when each pair has at most one feasible agreement, NOM holds for every stable-dominating mechanism.

\begin{theorem}\label{thm:unique-contracts}
In the many-to-many matching model with unitary contracts, suppose that
\[
|X_d\cap X_h|\leq1\qquad\text{for every }d\in D\text{ and }h\in H.
\]
Then every stable-dominating mechanism, including every stable mechanism and every efficient stable-dominating mechanism, is NOM.
\end{theorem}

\begin{proof}
Fix a doctor $d$ and a true preference $P_d$. Define
\[
G_d=\{x\in X_d:x\in C_{x_H}(X_{x_H})\},
\qquad
L=C_d(P_d,G_d).
\]
By substitutability, each contract in $G_d$ is chosen by
its hospital from every offer set containing it.

\medskip
\noindent\textbf{Claim 1.}
Every stable-dominating mechanism $\varphi$ satisfies
\[
\varphi_d(P_d,P_{-d})R_dL
\qquad\text{for every }P_{-d}.
\]

\smallskip
\noindent\emph{Proof of Claim 1.}
It suffices to establish the inequality for every stable
allocation $Y$ at $(P_d,P_{-d})$.
Suppose instead that $LP_dY_d$. Since
$L=C_d(P_d,G_d)$, there exists
\[
x\in C_d(P_d,Y_d\cup G_d)\setminus Y_d.
\]
In particular, $x\in G_d$. Substitutability implies
\[
x\in C_d(P_d,Y_d\cup\{x\}),
\qquad
x\in C_{x_H}(Y_{x_H}\cup\{x\}).
\]
Thus $x$ blocks $Y$, a contradiction.
The conclusion follows from stable domination. \hfill$\square$

\medskip
\noindent\textbf{Claim 2.}
There exists a subprofile $\widehat P_{-d}$ such that,
for every declaration $P'_d$ and every stable-dominating
mechanism $\varphi$,
\[
\varphi_d(P'_d,\widehat P_{-d})=C_d(P'_d,G_d).
\]

\smallskip
\noindent\emph{Proof of Claim 2.}
For each hospital $h$, define
\[
B_h=
\begin{cases}
\varnothing,
& \text{if }X_d\cap X_h\cap G_d\ne\varnothing,\\
C_h(X_h),
& \text{otherwise},
\end{cases}
\qquad
B=\bigcup_{h\in H}B_h.
\]
Since each doctor--hospital pair has at most one feasible
contract, $B\cap X_d=\varnothing$.

Let $\widehat P_i$, for each $i\ne d$, be a strict,
additively separable preference under which precisely
the contracts in $B_i$ have positive utility.
These preferences are substitutable, and
\[
C_i(\widehat P_i,Y)=Y_i\cap B_i
\qquad\text{for every }Y\subseteq X.
\]

Fix $P'_d$. Every stable allocation at
$(P'_d,\widehat P_{-d})$ must contain $B$.
Indeed, each $b\in B_h$ belongs to $C_h(X_h)$, so
substitutability implies that its hospital accepts it
from every offer set containing it. Its doctor also
accepts it whenever it is available. Hence any missing
contract in $B$ would block the allocation.

If $x\in X_d\setminus G_d$ and $h=x_H$, then
$B_h=C_h(X_h)$ and $x\notin B_h$. Choice consistency gives
\[
C_h(B_h\cup\{x\})=B_h.
\]
Individual rationality therefore excludes $x$ from any
allocation containing $B$: the other doctors accept
only contracts in $B$, and $x$ is the only contract
between $d$ and $h$.

Consequently, every stable allocation contains $B$ and
assigns $d$ a subset of $G_d$. Consider
\[
S'=B\cup C_d(P'_d,G_d).
\]
This allocation is individually rational. At hospitals
with $B_h\ne\varnothing$, the assigned contracts are
exactly $B_h=C_h(X_h)$. At the remaining hospitals,
the assignment is either empty or a singleton belonging
to $G_d$. Individual rationality for doctors follows
from their construction and choice consistency.
There is no blocking contract involving another doctor,
since all contracts acceptable to these doctors already
belong to $B$. A contract in $X_d\setminus G_d$ is rejected
by its hospital when added to $S'$, because
$C_h(B_h\cup\{x\})=B_h$. A contract in
$G_d\setminus C_d(P'_d,G_d)$ is rejected by $d$ when
added to its assignment, by choice consistency.
Thus $S'$ is stable.

To prove uniqueness, let $Y$ be another stable allocation. Then
$Y_{-d}=B$ and $Y_d\subseteq G_d$. If
$Y_d\ne C_d(P'_d,G_d)$, there exists
\[
x\in C_d(P'_d,Y_d\cup G_d)\setminus Y_d.
\]
Substitutability implies that $d$ accepts $x$ from
$Y_d\cup\{x\}$, and its hospital accepts it because
$x\in G_d$. Thus $x$ blocks $Y$, a contradiction.

Finally, any individually rational allocation weakly
Pareto-dominating $S'$ must retain $B$, since $B_i$ is
the most preferred assignment of each doctor $i\ne d$.
It therefore assigns $d$ a subset of $G_d$.
As $C_d(P'_d,G_d)$ is the most preferred such subset,
that allocation must coincide with $S'$.
Stable domination proves the claim. \hfill$\square$

Claims 1 and 2 imply
\[
W_d(P_d,O^\varphi(P_d))=L.
\]
For every alternative declaration $P'_d$, Claim 2 gives
\[
\begin{aligned}
L
&\;R_d\;C_d(P'_d,G_d)\\
&=\varphi_d(P'_d,\widehat P_{-d})\\
&\;R_d\;W_d(P_d,O^\varphi(P'_d)).
\end{aligned}
\]
Thus $\varphi$ admits no obvious manipulation in the
worst case.

\medskip
\noindent\textbf{Claim 3.}
Let
\[
E_d=\{x\in X_d:C_{x_H}(\{x\})=\{x\}\},
\qquad
Z=C_d(P_d,E_d).
\]
For every stable-dominating mechanism $\varphi$,
\[
B_d(P_d,O^\varphi(P_d))=Z,
\]
and no outcome under any alternative declaration is
preferred to $Z$ according to $P_d$.

\smallskip
\noindent\emph{Proof of Claim 3.}
Let every other doctor accept no contracts. Since
each doctor--hospital pair has at most one feasible
contract, $Z$ is individually rational and stable.

Moreover, it is the unique stable allocation at this
profile. Indeed, any individually rational allocation
$Y$ assigns nothing to the other doctors and satisfies
$Y_d\subseteq E_d$. If $Y_d\ne Z$, there exists
\[
x\in C_d(P_d,Y_d\cup E_d)\setminus Y_d.
\]
By substitutability, $d$ accepts $x$ from
$Y_d\cup\{x\}$. Its hospital is unmatched at $Y$,
because there is no other contract between $d$ and
$x_H$. Since $x\in E_d$, this contract blocks $Y$.

Every individually rational allocation at this profile
assigns $d$ a subset of $E_d$, so none can make $d$
strictly better off than $Z$. Stable domination
therefore requires $\varphi$ to select $Z$.

Finally, hospital individual rationality and
substitutability imply that $d$ receives a subset of
$E_d$ under every declaration and every subprofile.
Since $Z=C_d(P_d,E_d)$, no such outcome is preferred
to $Z$ according to $P_d$.\hfill$\square$

Claim 3 rules out obvious manipulations in the best
case. Together with the worst-case comparison, this
proves that $\varphi$ is NOM.
\end{proof}

\begin{remark}\label{rem:no-contracts}
The restriction in Theorem~\ref{thm:unique-contracts} makes every preference unitary, so $\mathcal P^u=\mathcal P$. Thus the result covers the full substitutable domain in these markets. With a unique feasible agreement for each possible pair, the contractual structure can be interpreted as the classical many-to-many matching model without contracts. The theorem extends the many-to-one benchmark of \citet{troyan2020obvious,arribillaga2025obvious} to doctors who may receive sets of contracts. Its conclusion concerns every stable-dominating mechanism, without requiring efficiency.
\end{remark}

\section{Final Remarks}\label{sec:final}
Table~\ref{tab:results} summarizes the comparison between the full substitutable domain and the unitary domain.

\begin{table}[ht]
\centering\small
\setlength{\tabcolsep}{3pt}
\begin{tabular}{>{\raggedright\arraybackslash}p{.32\linewidth}>{\centering\arraybackslash}p{.32\linewidth}>{\centering\arraybackslash}p{.30\linewidth}}
\toprule
Mechanism/class & Full substitutable domain & Unitary domain\\
\midrule
$DDAM$ & OM in both cases & NOM\\
$HDAM$ & OM in both cases & OM in both cases\\
Stable mechanisms & OM in both cases & Depends on the mechanism\\
Efficient stable-dominating & OM in the worst case only & OM in the worst case only\\
\bottomrule
\end{tabular}
\caption{Obvious manipulation in the worst case and in the best case.}\label{tab:results}

\smallskip
\begin{minipage}{.97\linewidth}
\footnotesize
OM entries refer to markets satisfying the sufficient conditions in Section~\ref{sec:competition}; NOM holds in every admissible market on the stated domain. If each doctor--hospital pair has at most one feasible contract, every stable-dominating mechanism is NOM by Theorem~\ref{thm:unique-contracts}.
\end{minipage}
\end{table}

The comparison gives two distinct lessons. First, the contractual structure and the unitary preference restriction affect which stable mechanisms remain NOM. Second, efficiency has a specific strategic implication: it prevents best-case obvious manipulation but need not prevent a worst-case improvement. The competing, protected-competing, and replacement conditions identify sufficient choice configurations behind these conclusions. They do not provide necessary conditions, and their failure does not by itself establish NOM.

The results leave open the behavior of stable-dominating mechanisms that are neither stable nor efficient in markets with alternative contracts. They also leave open a characterization of NOM stable mechanisms on the unitary domain. In addition, our comparison takes the available contract set as given; it does not establish that changing the contractual representation of a fixed application preserves preferences, stability, or strategic behavior. Exploring that connection would link the present analysis more directly to endogenous contract design.

All results use the stated information structure, with fixed hospital preferences. Substitutability also matters for the existence and choice properties used in the proofs. The following example describes behavior outside that assumption on a restricted domain where the mechanism is well defined. Since it allows alternative contracts for a pair, it is not a counterexample to Theorem~\ref{thm:unique-contracts}, nor does it imply that every violation of substitutability produces obvious manipulation.

\begin{example}\label{ex:nonsub}
Let $D=\{d_1,d_2\}$, $H=\{h\}$, and $X=\{x,y,z\}$, where $x,y$ involve $d_1$, $z$ involves $d_2$, and all contracts involve $h$. Let
\[
 P_h:\{y,z\},\{x\},\{y\},\{z\},\varnothing,
 \quad P_1:\{x,y\},\{x\},\{y\},\varnothing,
 \quad P_2:\{z\},\varnothing.
\]
At this profile the unique stable and efficient stable-dominating allocation is $\{y,z\}$. Under $P'_1:\{x\},\varnothing$, it is $\{x\}$. Under $P_1$ with $d_2$ rejecting $z$, the unique stable allocation is also $\{x\}$. On the product subdomain consisting of these two reports for each doctor, the resulting mechanism therefore has option sets $\{\{y\},\{x\}\}$ under $P_1$ and $\{\{x\}\}$ under $P'_1$. The worst case improves, while the best case is unchanged. This is an obvious manipulation in the worst case on a domain where the relevant mechanisms are well defined, with nonsubstitutable hospital preferences.
\end{example}

\bibliographystyle{ecta}
\bibliography{biblio}

@article{ortega2022obvious,
  title={Obvious manipulations in cake-cutting},
  author={Ortega, Josu{\'e} and Segal-Halevi, Erel},
  journal={Social Choice and Welfare},
  pages={1--20},
  year={2022},
  publisher={Springer}
}

@article{psomas2022fair,
  title={Fair and efficient allocations without obvious manipulations},
  author={Psomas, Alexandros and Verma, Paritosh},
  journal={arXiv preprint arXiv:2206.11143},
  year={2022}
}

@inproceedings{aziz2021obvious,
  title={Obvious manipulability of voting rules},
  author={Aziz, Haris and Lam, Alexander},
  booktitle={International Conference on Algorithmic Decision Theory},
  pages={179--193},
  year={2021},
  organization={Springer}
}

@article{troyan2020obvious,
  title={Obvious manipulations},
  author={Troyan, Peter and Morrill, Thayer},
  journal={Journal of Economic Theory},
  volume={185},
  pages={104970},
  year={2020},
  publisher={Elsevier}
}

@article{echenique2004theory,
  title={A theory of stability in many-to-many matching markets},
  author={Echenique, F. and Oviedo, J.},
  journal={Theoretical Economics},
  volume={1},
  number={1},
  pages={233--273},
  year={2006},
  publisher={}
}

@article{kesten2010school,
  title={School choice with consent},
  author={Kesten, Onur},
  journal={The Quarterly Journal of Economics},
  volume={125},
  number={3},
  pages={1297--1348},
  year={2010},
  publisher={MIT Press}
}

@article{alva2019strategy,
  title={Strategy-proof Pareto-improvement},
  author={Alva, Samson and Manjunath, Vikram},
  journal={Journal of Economic Theory},
  volume={181},
  pages={121--142},
  year={2019},
  publisher={Elsevier}
}

@article{chen2023regret,
  title={Regret-free truth-telling in school choice with consent},
  author={Chen, Yiqiu and M{\"o}ller, Markus},
  journal={Theoretical Economics},
  volume={19},
  pages={635--666},
  year={2024},
  doi={10.3982/TE4979}
}

@article{arribillaga2024obvious,
  title={Obvious manipulations of tops-only voting rules},
  author={Arribillaga, R Pablo and Bonifacio, Agust{\'\i}n G},
  journal={Games and Economic Behavior},
  volume={143},
  pages={12--24},
  year={2024},
  publisher={Elsevier}
}

@article{kominers2012correspondence,
  title={On the correspondence of contracts to salaries in (many-to-many) matching},
  author={Kominers, Scott Duke},
  journal={Games and Economic Behavior},
  volume={75},
  number={2},
  pages={984--989},
  year={2012},
  publisher={Elsevier}
}

@article{hatfield2017contract,
  title={Contract design and stability in many-to-many matching},
  author={Hatfield, John William and Kominers, Scott Duke},
  journal={Games and Economic Behavior},
  volume={101},
  pages={78--97},
  year={2017},
  publisher={Elsevier}
}

@article{klaus2009stable,
  title={Stable many-to-many matchings with contracts},
  author={Klaus, Bettina and Walzl, Markus},
  journal={Journal of Mathematical Economics},
  volume={45},
  number={7-8},
  pages={422--434},
  year={2009},
  publisher={Elsevier}
}

@article{dougan2021minimally,
  title={Minimally unstable Pareto improvements over deferred acceptance},
  author={Do{\u{g}}an, Battal and Ehlers, Lars},
  journal={Theoretical Economics},
  volume={16},
  number={4},
  pages={1249--1279},
  year={2021},
  publisher={Wiley Online Library}
}

@article{arribillaga2025obvious,
  title={Obvious manipulations in matching with and without contracts},
  author={Arribillaga, R Pablo and Pepa Risma, Eliana},
  journal={Games and Economic Behavior},
  volume={151},
  pages={70--81},
  year={2025},
  publisher={Elsevier}
}

@article{hatfield2014many,
  title={Many-to-many matching with max--min preferences},
  author={Hatfield, John William and Kojima, Fuhito and Narita, Yusuke},
  journal={Discrete Applied Mathematics},
  volume={179},
  pages={235--240},
  year={2014},
  publisher={Elsevier}
}

@article{chen2016manipulability,
  title={Manipulability of stable mechanisms},
  author={Chen, Peter and Egesdal, Michael and Pycia, Marek and Yenmez, M Bumin},
  journal={American Economic Journal: Microeconomics},
  volume={8},
  number={2},
  pages={202--214},
  year={2016},
  publisher={American Economic Association 2014 Broadway, Suite 305, Nashville, TN 37203-2425}
}

@article{Blair1988,
  author  = {Blair, Charles},
  title   = {The Lattice Structure of the Set of Stable Matchings with Multiple Partners},
  journal = {Mathematics of Operations Research},
  volume  = {13},
  number  = {4},
  pages   = {619--628},
  year    = {1988}
}

@unpublished{alva2019stabledominating,
  author = {Alva, Samson and Manjunath, Vikram},
  title = {Stable-dominating rules},
  year = {2019},
  note = {Working paper, September 4},
  url = {https://samsonalva.com/StableDomination.pdf}
}

@techreport{makominers2018bundling,
  author = {Ma, Jonathan and Kominers, Scott Duke},
  title = {Bundling Incentives in (Many-to-Many) Matching with Contracts},
  institution = {Harvard Business School},
  type = {Working Paper}, number = {19-011}, year = {2018}
}

@article{archbold2025singleparameter,
  author = {Archbold, Thomas and de Keijzer, Bart and Ventre, Carmine},
  title = {Non-Obvious Manipulability for Single-Parameter Agents and Bilateral Trade},
  journal = {ACM Transactions on Economics and Computation},
  year = {2025}, volume = {13}, number = {4},
  pages = {16:1--16:26}, doi = {10.1145/3746458}
}

@article{dekeijzer2026budget,
  author = {de Keijzer, Bart and Sch{\"a}fer, Guido and Tsikiridis, Artem and Ventre, Carmine},
  title = {Breaking Barriers, Finding Boundaries: Not Obviously Manipulable Budget-Feasible Mechanism Design},
  journal = {Proceedings of the AAAI Conference on Artificial Intelligence},
  year = {2026}, volume = {40}, number = {20}, pages = {16803--16811},
  doi = {10.1609/aaai.v40i20.38724}
}

@unpublished{chen2026disclosure,
  author = {Chen, Siwei},
  title  = {Information Disclosure and Incentive Collapse
            in Division Problems},
  year   = {2026},
  note   = {Mimeo}
}

@article{article,
author = {Arribillaga, R. Pablo and Bonifacio, Agustín G.},
year = {2025},
month = {01},
pages = {355-380},
title = {Not obviously manipulable allotment rules},
volume = {80},
journal = {Economic Theory},
doi = {10.1007/s00199-024-01633-1}
}

@article{unknown,
  author  = {Arribillaga, R. Pablo and Bonifacio, Agustín G.},
  title   = {Obvious manipulations, consistency, and the uniform rule},
  journal = {Economics Letters},
  year    = {2025},
  volume  = {252},
  pages   = {112344},
  doi     = {10.1016/j.econlet.2025.112344}
}

\end{document}